\documentclass[10pt,a4paper]{article}

\usepackage{caption}
\usepackage{subcaption}
\usepackage{graphicx}
\usepackage{float}
\graphicspath{{figures/}}

\usepackage{amssymb}
\usepackage{amsmath}
\usepackage{amsthm}
\usepackage{mathtools}

\newtheorem{proposition}{Proposition}[section]

\newtheorem{assumption}[proposition]{Assumption}
\theoremstyle{remark}

\usepackage{algorithm}
\usepackage{algorithmic}
\usepackage{array}
\usepackage{booktabs}
\usepackage{hyperref}
\usepackage[english]{babel}
\usepackage{enumitem}
\usepackage{xcolor}
\usepackage{listings}
\usepackage{microtype}
\usepackage{siunitx}

\hypersetup{
    colorlinks=true,
    linkcolor=blue,
    citecolor=blue,
    urlcolor=blue
}

\makeatletter
\renewcommand\@maketitle{%
  \newpage
  \null
  \vskip 2em%
  \begin{center}%
  \let\footnote\thanks
    {\LARGE\@title\par}%
    \vskip 1.5em%
    {\normalsize
      \lineskip .5em%
      \begin{tabular}[t]{c}%
        \@author
      \end{tabular}\par}%
  \end{center}%
  \par
  \vskip 1.5em}
\makeatother

\begin{document}

\title{Redefining Hybrid Blockchains: A Balanced Architecture}
\author{\textbf{Syed Ibrahim Omer}\\City University of Hong Kong\\\href{https://github.com/ibitec7/blockchain}{github.com/ibitec7/blockchain}}
\date{}
\maketitle

\begin{abstract}
\vspace{0.5em}
\begin{center}
\begin{minipage}{0.79\textwidth}
\normalsize
Blockchain technology has completely revolutionized the field of
decentralized finance with the emergence of a variety of cryptocurrencies and
digital assets. However, widespread adoption of this technology by governments
and enterprises has been limited by concerns regarding the technology's
scalability, governance, and economic sustainability. This paper aims to introduce
a novel hybrid blockchain architecture that balances scalability, governance, and
decentralization while being economically viable for all parties involved. The new
semi-centralized model leverages strategies not prevalent in the field, such as
resource and node isolation, containerization, separation of networking and
compute layers, use of a Kafka pub-sub network instead of a peer-to-peer
network, and stakes-based validator selection to possibly mitigate a variety of
issues related to scalability, security, governance, and economic sustainability.
Simulations conducted on Kubernetes demonstrate the architecture's ability to
achieve over 1000 transactions per second, with consistent performance across
scaled deployments, even on a lightweight consumer-grade laptop with resource
constraints. The findings highlight the system's scalability, security, and economic
viability, offering a robust framework for enterprise and government adoption.
\end{minipage}
\end{center}
\end{abstract}

\section{Introduction}
\label{sec:introduction}

Blockchain technology changed how distributed systems establish trust: a network of validators that agree on a single state gives data integrity and fault tolerance without a central authority. Governments and enterprises have nevertheless been slow to adopt it, and three concerns dominate their hesitation: scalability, governance, and economic sustainability \cite{zheng2018blockchain,bonneau2015sok}.

The governance concern is the sharpest. An organization holding sensitive data cannot let unvetted public validators influence it, so such organizations deploy private or permissioned chains that grant authority to the operator. This resolves governance but creates new problems: the operator must host, maintain, and scale the entire distributed system, and that cost grows with the network. Public chains avoid the burden but surrender governance, since authority rests with public participants. Permissioned chains keep governability but lose transparency and public trust, and their scalability depends on the operator's budget \cite{androulaki2018fabric}. Neither extreme satisfies an organization that needs governance alongside the integrity and transparency of a public chain.

A suitable architecture must be scalable, economical to operate, and must give the hosting organization effective governance over its data while still providing public verifiability. Public and permissioned chains each trade one of these properties for another, and the hybrid chains that combine them inherit unresolved problems where the two halves meet.

\subsection{Gaps in Existing Approaches}

We identify four gaps that motivate this work.

\begin{enumerate}[label=(\roman*), leftmargin=1.7em, itemsep=1pt, topsep=3pt]
    \item \textbf{Governance or decentralization, not both.} Public chains decentralize consensus but remove organizational control, while permissioned chains do the opposite. No mainstream design fixes the degree of centralization to a known, bounded quantity.
    \item \textbf{Cost of scaling falls on one party.} In a permissioned deployment the hosting organization pays for the whole distributed infrastructure, and in a deployment that admits public validators the governance problem returns.
    \item \textbf{Weak security model.} Proof-of-Work and Proof-of-Stake secure the chain under economic assumptions a state-level adversary can violate, and the resulting centralization is unpredictable and hard to reverse once it forms \cite{eyal2014majority}.
    \item \textbf{Middleware used narrowly.} Kafka has been deployed almost exclusively as an ordering service that sequences transactions \cite{androulaki2018fabric}. Its use as the general communication substrate of a blockchain, or the use of containerized microservices to structure chain components, has received little attention \cite{dragoni2017microservices}.
\end{enumerate}

\subsection{Contributions}

This paper makes the following contributions.

\begin{enumerate}[label=(\roman*), leftmargin=1.7em, itemsep=1pt, topsep=3pt]
    \item \textbf{A semi-centralized architecture} that separates the consensus protocol from the network protocol. Consensus remains decentralized and is executed by public validators; the network protocol is governed by a single master node and fixes the degree of centralization to a constant, known quantity.
    \item \textbf{Kafka as a communication substrate.} Nodes communicate through a Kafka pub-sub middleware rather than directly, which isolates nodes, removes direct peer-to-peer attack channels, and lets the network layer scale independently of consensus.
    \item \textbf{A stake-weighted validator selection policy} implemented by the master node, which admits validators in proportion to their stake and can interrupt monopolies that would otherwise form.
    \item \textbf{A containerized, Kubernetes-native deployment} in which every role runs as an independent microservice with its own resource limits, giving fault isolation and independent scaling.
    \item \textbf{An empirical study of the deployed system.} We measure throughput, finality, and synchronization across single-node, three-node, and four-node deployments on consumer-grade hardware, and report where the system meets and misses its targets.
\end{enumerate}

\subsection{Organization}

Section~\ref{sec:related-work} reviews related work. Section~\ref{sec:method} presents the system model, the architecture, and its implementation. Section~\ref{sec:results} reports the empirical evaluation, and Section~\ref{sec:conclusion} concludes. Detailed derivations, the extended discussion, and configuration tables appear in the appendix.

\section{Related Work}
\label{sec:related-work}

This work sits at the meeting point of four research threads: blockchain architecture and governance, consensus mechanisms, scalability techniques, and the distributed middleware and deployment models that carry the chain's traffic.

\subsection{Blockchain Architectures and Governance}

Blockchains fall into three broad classes that differ in who may validate and who governs. Public chains admit any participant and build trust through collective consensus, most recently surveyed by Bonneau et al.\ \cite{bonneau2015sok} and Zheng et al.\ \cite{zheng2018blockchain}. Permissioned chains restrict participation to approved parties and thereby grant the hosting organization control over the data \cite{androulaki2018fabric}. Hybrid chains combine elements of both to obtain a middle policy \cite{sagirlar2018hybridiot}. The middle policy is the difficult one: most hybrid designs either loosen a permissioned chain by admitting a trusted consortium, which does not grant public validators any real role, or they connect a private chain to a public chain for anchoring, which introduces an integration boundary.

\subsection{Consensus Mechanisms}

Proof-of-Work secures Bitcoin through the cryptographic work committed to the chain \cite{nakamoto2008bitcoin}. Its cost is throughput and energy: the protocol targets one block per ten minutes and global Bitcoin energy use is on the order of 175\,TWh per year \cite{statista2024bitcoin,pakistan2024energy,vranken2017sustainability}. The security argument holds only while honest nodes control a majority of the hash rate, and majority control is not sufficient for correctness, since rational miners can profit from deviating from the protocol \cite{eyal2014majority}. Proof-of-Stake selects block proposers in proportion to their stake, so a dishonest actor risks capital \cite{buterin2017casper,lau2020ethereum}. The stake requirement is folded into the security argument, and it holds only while acquiring a majority stake is impractical. State-level actors may not face that constraint, and stake concentrates over time in a way that is unpredictable and difficult to reverse.

Byzantine fault-tolerant protocols offer a third path. PBFT provides safety and liveness with $3f+1$ replicas and quadratic message complexity \cite{castro1999pbft}. Later protocols reduce the communication cost, including Tendermint \cite{buchman2018tendermint}, Algorand \cite{gilad2017algorand}, HoneyBadgerBFT \cite{miller2016honeybadger}, and HotStuff \cite{yin2019hotstuff}. These protocols assume a bounded number of faulty validators rather than an economic majority. They are the natural fit for a system that must give the hosting organization control over validator composition, and we build on this family.

\subsection{Scalability Techniques}

Throughput limits in early chains motivated three families of techniques. Layer-two proposals move work off the main chain: the Lightning Network settles micropayments off-chain \cite{poon2016lightning}, and rollups and state channels post compressed results or proofs to the chain \cite{ethereum2024scaling}. These approaches rely on localized trust and leave governance to the base layer. Sharding partitions state and workload across smaller chains for parallel processing \cite{ethereum2024scaling}, but cross-shard communication adds synchronization overhead and complexity. Consensus redesign, the family we follow, attacks the limit at the protocol level. Tendermint separates consensus from networking so the two can scale independently \cite{buchman2018tendermint}, a separation we adopt.

\subsection{Ordering, Messaging, and Deployment Middleware}

Apache Kafka is a distributed log built for high-throughput, real-time data pipelines \cite{kreps2011kafka}. It has been used in blockchains chiefly as an ordering service that batches and sequences transactions before they are processed, a role it plays in Hyperledger Fabric \cite{androulaki2018fabric}. This uses only a fraction of Kafka's capability. Kafka can also act as the general transport between validators, providing store-and-replay streams, replication, and failover, none of which the ordering-service role exploits. Container orchestration systems derived from Borg, Omega, and Kubernetes schedule and isolate services across a cluster \cite{burns2016borg}, and the microservice pattern structures an application as small, independently deployable services \cite{dragoni2017microservices}. Blockchains have adopted microservices only in narrow settings, for example identity services in Hyperledger Indy \cite{indy2024}. A full chain built as containerized microservices has not been studied.

\subsection{Research Gap}

Table~\ref{tab:gap} compares representative architectures against the properties an enterprise deployment needs. Public and permissioned chains each satisfy a disjoint subset. Hybrid designs improve on one axis and regress on another. No design in the table fixes the level of centralization to a constant, uses middleware for all inter-validator communication, and packages the chain as independently scalable services. This paper targets exactly that combination.

\begin{table}[!htbp]
\centering
\caption{Representative architectures against the properties an enterprise deployment requires. A filled cell marks a property the design provides by construction.}
\label{tab:gap}
\small
\begin{tabular*}{\textwidth}{@{\extracolsep{\fill}}l c c c c c@{}}
\toprule
\textbf{Architecture} & \textbf{Bounded} & \textbf{Public} & \textbf{Independent} & \textbf{Isolated} & \textbf{Fault} \\
 & \textbf{centralization} & \textbf{validators} & \textbf{scaling} & \textbf{network} & \textbf{isolation} \\
\midrule
Public chains \cite{nakamoto2008bitcoin}      &          & \checkmark &          &          &          \\
Permissioned chains \cite{androulaki2018fabric} & \checkmark &            &          &          &          \\
Layer-two \cite{poon2016lightning}            &          &            & \checkmark &          &          \\
Sharded chains \cite{kang2018chainspace}      &          & \checkmark & \checkmark &          &          \\
Hybrid-IoT \cite{sagirlar2018hybridiot}       & \checkmark &            &          &          &          \\
This work                                     & \checkmark & \checkmark & \checkmark & \checkmark & \checkmark \\
\bottomrule
\end{tabular*}
\end{table}

\section{Method}
\label{sec:method}

This section specifies the system model, the architecture, the consensus and selection workflow, and the implementation. We use the following notation. Let $n$ be the number of validator nodes in the network, and let $b$ be the number of Kafka brokers. A validator $i$ holds stake $s_i$ and a public key. A round is one complete consensus instance that commits one block. A block holds at most $B$ transactions.

\subsection{Preliminaries}

\paragraph{Byzantine fault tolerance.} We adopt PBFT because it tolerates up to $f$ Byzantine validators among $3f+1$ and exchanges messages between validators, which stresses the communication layer we want to measure \cite{castro1999pbft}. The protocol runs in a pre-prepare, prepare, and commit sequence, and we treat it as a replaceable component.

\paragraph{Kafka.} Kafka is a distributed log where producers append records to topics and consumers read them in order \cite{kreps2011kafka}. We use Kafka as the message transport between nodes rather than as an ordering service, so all traffic passes through a replicated log instead of direct links.

\paragraph{Kubernetes.} Kubernetes schedules containerized services as pods, and each pod has resource requests and limits that isolate it from its neighbours \cite{burns2016borg}. Every role in our architecture is a separate pod.

\subsection{System Model and Trust Assumptions}

The system has two governing roles and one transport role.

\begin{description}[leftmargin=*, nosep]
    \item[Validator nodes] process transactions, build blocks, and run the consensus protocol. They are operated by parties other than the hosting organization and may be untrusted individually.
    \item[Master node] selects the validators that participate in each round and governs the network protocol. It is operated by the hosting organization and is trusted for liveness and for the fairness of selection.
    \item[Kafka brokers] carry all messages between validators, the master, and clients. They are operated by the hosting organization.
\end{description}

\begin{assumption}[Trust model]
\label{ass:trust}
The master node and the Kafka brokers are honest but may fail. The consensus protocol tolerates up to $f$ Byzantine validators. Clients are untrusted. The hosting organization accepts the master node as the sole authority over validator composition.
\end{assumption}

The model is deliberately narrow. It does not claim that the network is decentralized; it fixes centralization to the master node and the brokers, both of which are known to participants in advance. This is a weaker trust assumption than a public chain places on its validator majority, and a stronger one than a permissioned chain places on its operator alone, because consensus remains open to external validators.

\subsection{Architecture Overview}

The architecture separates two protocols that mainstream chains fuse. The consensus protocol is decentralized and is executed by the validator nodes, which are responsible for data integrity. The network protocol is centralized and is executed by the master node, which is responsible for who may communicate and who may be selected. Kafka brokers sit between the two and carry every message, so no validator communicates with another directly.

\begin{figure}[!htbp]
\centering
\begin{subfigure}[t]{0.43\textwidth}
    \centering
    \includegraphics[width=\textwidth]{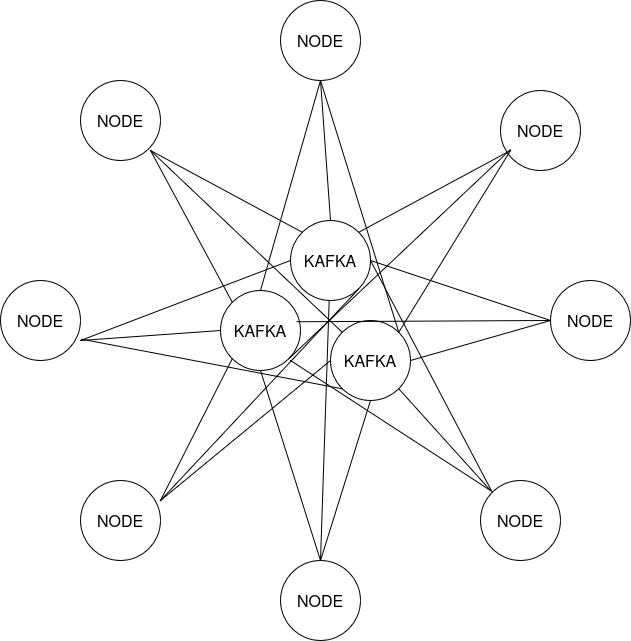}
    \caption{The proposed architecture. Validators and the master exchange messages through Kafka brokers, so no direct links exist between validators.}
    \label{fig:arch}
\end{subfigure}
\hfill
\begin{subfigure}[t]{0.43\textwidth}
    \centering
    \includegraphics[width=\textwidth]{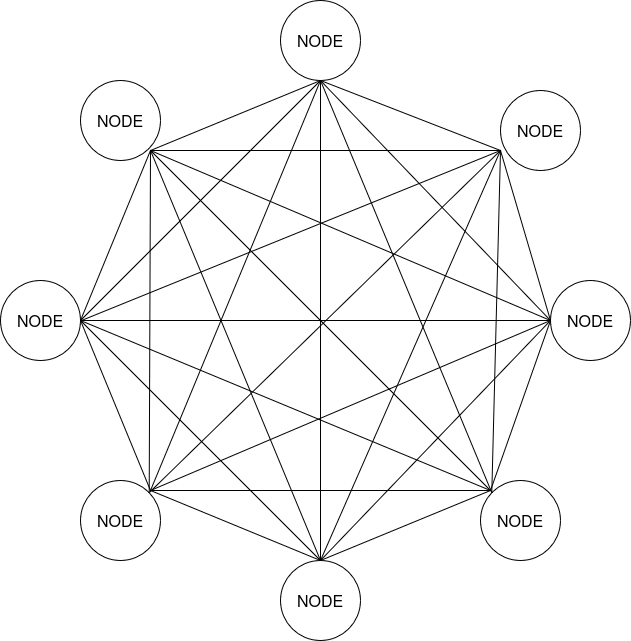}
    \caption{A peer-to-peer network, where every validator connects to every other validator.}
    \label{fig:p2p}
\end{subfigure}
\caption{Communication topology. The proposed architecture routes all traffic through brokers (left); a peer-to-peer network creates a direct link between every pair of validators (right).}
\label{fig:topology}
\end{figure}

Because the master node mediates selection, it can exclude a validator that has accumulated enough stake to control a round. Consensus itself remains open to validators the master has not excluded, so data integrity rests on the decentralized half of the system while composition rests on the centralized half.

\subsection{Kafka-Mediated Communication}

In a peer-to-peer network each validator maintains a connection to every other validator, as shown in Figure~\ref{fig:topology}b. Every message is delivered directly, so a malicious validator holds a direct channel to every honest validator. The architecture replaces these links with a Kafka topic that all parties publish to and subscribe to, shown in Figure~\ref{fig:topology}a. Messages are appended to partitions and read in order, and a validator that fails to authenticate with the master is not admitted to the topic. This removes the direct channel that Sybil and message-injection attacks use \cite{douceur2002sybil}. Kafka's partition replication also absorbs a broker failure without interrupting delivery, and the per-phase consumer offsets give every validator the same view of the round.

\begin{figure}[!htbp]
\centering
\includegraphics[width=0.58\textwidth]{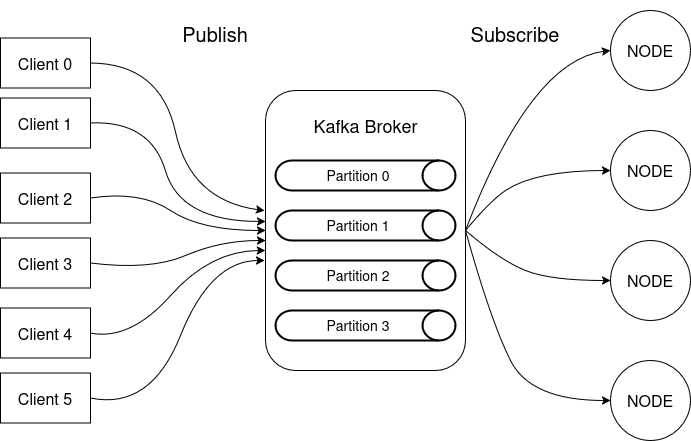}
\caption{The pub-sub model used for transaction and consensus messaging. Clients publish to the broker, which distributes records across partitions to subscribed validators.}
\label{fig:pubsub}
\end{figure}

\subsection{Workflow}

A round proceeds in three phases: transaction flow, validator selection, and consensus.

\paragraph{Transaction flow.} Clients submit transactions either directly to a validator or through a dedicated Kafka topic. The topic path supports queueing and ordering and is the intended mode for large deployments.

\paragraph{Validator selection.} When validators have prepared a block, they signal readiness to the master and propose their stake for the round. The master draws a subset of validators using a verifiable random function weighted by each validator's share of stake. The draw is verifiable, so a validator can confirm that it was selected fairly without trusting the master's computation.

\paragraph{Consensus.} Selected validators execute PBFT through the Kafka topic. Participants run the protocol exactly as they would in a peer-to-peer deployment, with the difference that messages pass through the master and the brokers. The master also synchronizes the round: it holds the start of each phase until every participant is ready, which removes the skew that causes divergent block timestamps.

\subsection{Governance and Fault Handling}

The level of centralization is constant by construction. The only centralized components are the master node and the brokers, and both are known to all participants. In Proof-of-Work and Proof-of-Stake the level of centralization emerges from the distribution of hash rate or stake and can drift without warning; here it cannot. The master cannot influence the content of blocks or the outcome of consensus, because it neither signs blocks nor votes. It controls only admission and scheduling.

If the master or the brokers fail, validators lose the ability to progress in coordination but not the ability to operate. Validators can continue to build on the last committed block and may temporarily fork; when the middleware recovers, consensus runs over the competing branches and reconciles them, which preserves both data integrity and progress.

\subsection{Complexity}

The network cost of the two topologies differs in how it grows with $n$.

\begin{proposition}[Network complexity]
\label{prop:complexity}
A peer-to-peer network among $n$ validators requires $\Theta(n^2)$ inter-node links, whereas the proposed architecture requires $\Theta(n)$ links between validators and brokers and $\Theta(b)$ links among brokers. In total the proposed architecture carries $O(n + b)$ links, independent of the pairing structure.
\end{proposition}

\begin{proof}
In a peer-to-peer network every validator connects to every other, which gives $\binom{n}{2} = \Theta(n^2)$ undirected links. In the proposed architecture each validator connects only to the broker set, giving at most $n \cdot b$ validator-to-broker links, and the brokers connect among themselves in $O(b)$ links. When $b$ is held constant, as in our deployment, both terms are linear in $n$. \end{proof}

\begin{figure}[!htbp]
\centering
\includegraphics[width=0.58\textwidth]{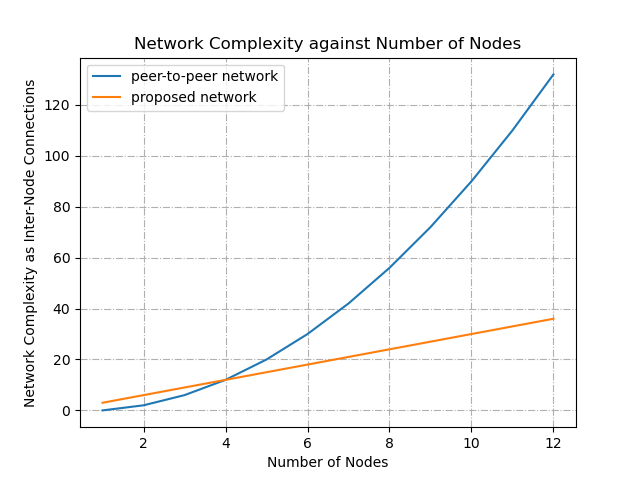}
\caption{Inter-node connections against the number of validators for a three-broker deployment. The peer-to-peer network grows quadratically; the proposed architecture grows linearly.}
\label{fig:complexity}
\end{figure}

The crossover favors the peer-to-peer network only while $n$ is small, and it moves in favour of the proposed architecture as the network grows. Figure~\ref{fig:complexity} plots both costs for a three-broker deployment; the proposed architecture overtakes the peer-to-peer network once the number of validators passes the number of brokers.

\subsection{Implementation}

The system is implemented in Rust. The language has no runtime or garbage collector, so the measured performance is not diluted by abstraction overhead, and its concurrency facilities support the asynchronous task model the nodes use.

Each role is a separate containerized service deployed on a Kubernetes cluster. \texttt{node}, \texttt{master}, and \texttt{transaction} pods run the validator, master, and transaction-generator roles, and Kafka and Zookeeper run as their own pods. Configuration is supplied through Kubernetes ConfigMaps, so block size, batch size, and topic settings are changed without rebuilding images.

\begin{figure}[!htbp]
\centering
\includegraphics[width=0.68\textwidth]{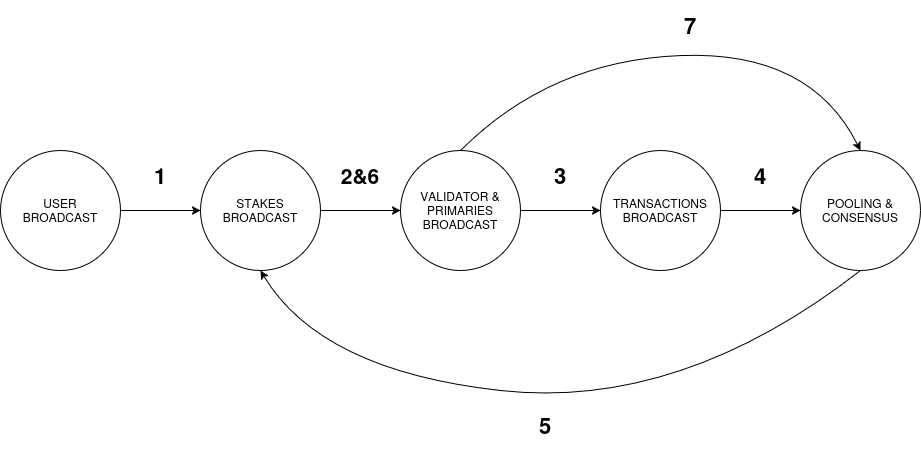}
\caption{The task automaton that schedules node execution. Each phase advances only after the preceding phase has broadcast its data, which removes the deadlock that arose from unordered asynchronous tasks.}
\label{fig:automaton}
\end{figure}

Transactions execute asynchronously because the number of simulated validators exceeds the number of available CPU threads. This introduces a runtime scheduling cost, discussed in Section~\ref{sec:results}. The node application runs its phases through a small state automaton that enforces ordering, shown in Figure~\ref{fig:automaton}; without it, a phase could run before the data it consumes is produced, which deadlocks the round.

We report the practical engineering steps that were required to make the deployment reproducible in Appendix~\ref{app:impl} and the resource and Kafka settings in Tables~\ref{tab:resources} and~\ref{tab:kafka}.

\section{Empirical Results}
\label{sec:results}

This section reports the evaluation of the deployed architecture. We state the setup, the metrics, and the measurements, and we interpret each result as it is introduced. The full interpretation, including the comparison with industry systems and the limits of the study, appears in Appendix~\ref{app:discussion}.

\subsection{Setup and Scope of Validation}

The evaluation measures scalability and performance. We do not claim to validate security, fault tolerance, or the economic model; testing those properties needs frameworks outside the scope of this deployment, and we return to them in Appendix~\ref{app:limitations}. Within the scope we fix a hypothesis: if the architecture scales as the complexity analysis predicts, throughput and finality should change little as validators are added.

The architecture runs on a local Kubernetes cluster on a single consumer laptop with a Ryzen 5 7535HS processor (6 cores, 12 threads, \SI{3.3}{\giga\hertz} base, \SI{4.6}{\giga\hertz} turbo) and \SI{32}{\giga\byte} of DDR5 memory at 4800\,MT/s. The cluster receives 10 CPU threads and \SI{30}{\giga\byte} of RAM. To make the measurement more representative of a real deployment, inter-service traffic is forced through the external Kafka port 9092 instead of the internal port 29092, which routes messages through the host network stack and adds the network overhead a deployed system would see.

The independent variable is the number of participating validators, held at one, three, and four nodes. Control variables, including per-node CPU and memory, network configuration, and the application binary, are held constant. The master node and the Kafka cluster are fixed across runs. The data set is 130 blocks of 512 transactions, or \num{66560} transactions, which is large enough for the distribution of each metric to be read directly. Each node times its own phases and appends the results to a CSV file.

\subsection{Metrics}

We report throughput at two levels. Transaction TPS counts individual transactions processed by the pool, and block TPS counts transactions committed through block creation and consensus. Block TPS reflects the network and consensus protocol introduced here, so it is the primary throughput metric. Latency is reported as consensus time, the time for one consensus round, and synchronization time, the time for nodes to reach a common starting state before a phase. Time to finality (TTF) is the maximum time from a transaction's submission to its inclusion in a committed block.

\subsection{Throughput and Latency}

Table~\ref{tab:stats} gives the five-number summary of the averaged metrics across the three deployments. The median block TPS is \num{1031.83}, and the median pool TPS is \num{407.45}. The median consensus time is \SI{496.28}{\milli\second}, of which synchronization accounts for \SI{43.01}{\milli\second}. The median total time per round is \SI{2263.51}{\milli\second} and the median TTF is \SI{2196.29}{\milli\second}.

Two features stand out. First, most of the round is spent pooling and processing transactions, about \SI{1.8}{\second} of the \SI{2.3}{\second} total, while consensus accounts for about \SI{0.5}{\second}. The consensus protocol itself is not the bottleneck; transaction preparation is. Second, synchronization is a small fraction of consensus time, at roughly \SI{43}{\milli\second} out of \SI{496}{\milli\second}, which indicates that Kafka carries the inter-node traffic without adding the delay that a peer-to-peer implementation might avoid. This supports the design choice of the middleware as a transport.

\begin{table}[!htbp]
\centering
\caption{Five-number summary of the averaged metrics across the three deployments. Times are in milliseconds.}
\label{tab:stats}
\small
\begin{tabular*}{\textwidth}{@{\extracolsep{\fill}}l r r r r r r r@{}}
\toprule
 & \textbf{Pool TPS} & \textbf{Block TPS} & \textbf{Pooling} & \textbf{Consensus} & \textbf{Sync} & \textbf{Total} & \textbf{TTF} \\
\midrule
Min    & 406.55 & 1016.45 & 1559.75 & 491.42 & 41.17 & 2063.44  & 1998.25 \\
Q1     & 407.07 & 1028.98 & 1566.42 & 495.17 & 42.37 & 2071.47  & 2005.63 \\
Median & 407.45 & 1031.83 & 1756.60 & 496.28 & 43.01 & 2263.51  & 2196.29 \\
Q3     & 407.87 & 1034.11 & 2591.76 & 497.69 & 43.93 & 3097.58  & 3030.37 \\
Max    & 409.39 & 1042.08 & 9824.61 & 504.06 & 51.81 & 10339.69 & 10271.36 \\
\bottomrule
\end{tabular*}
\end{table}

\subsection{Scalability Across Deployments}

Table~\ref{tab:delta} reports the percentage change in consensus metrics between consecutive deployments. Moving from one node to three changes block TPS by \SI{-1.83}{\percent} and consensus time by \SI{+1.89}{\percent}. Moving from three nodes to four changes block TPS by \SI{-1.10}{\percent} and consensus time by \SI{+1.12}{\percent}. Every consensus-phase metric other than commit time changes by less than about \SI{2}{\percent} across the full range.

The small changes confirm the scalability hypothesis. Adding validators does not degrade throughput or latency beyond a few percent, which the complexity analysis attributes to the linear link count of Proposition~\ref{prop:complexity} and the resource isolation of the containerized services. Commit time is the exception: it rises by \SI{77.36}{\percent} from single-node to three-node. The absolute value is small, an average of about \SI{13}{\milli\second}, so the large relative change does not affect the round total. We attribute the increase to the cost of aggregating votes from more nodes across Kafka.

\begin{table}[!htbp]
\centering
\caption{Percentage change in consensus metrics between consecutive deployments.}
\label{tab:delta}
\small
\begin{tabular*}{\textwidth}{@{\extracolsep{\fill}}l S[table-format=-1.2] S[table-format=+1.2] S[table-format=+1.2] S[table-format=+2.2] S[table-format=+1.2] S[table-format=+1.2]@{}}
\toprule
 & {\textbf{Block TPS}} & {\textbf{Pre-prepare}} & {\textbf{Prepare}} & {\textbf{Commit}} & {\textbf{Sync}} & {\textbf{Consensus}} \\
\midrule
Single $\to$ 3-node & -1.83 & +1.15 & +0.04 & +77.36 & +8.20 & +1.89 \\
3-node $\to$ 4-node & -1.10 & -0.15 & +0.39 & +25.48 & +1.57 & +1.12 \\
\bottomrule
\end{tabular*}
\end{table}

\subsection{Correlation Analysis}

\begin{figure}[!htbp]
\centering
\includegraphics[width=0.80\textwidth]{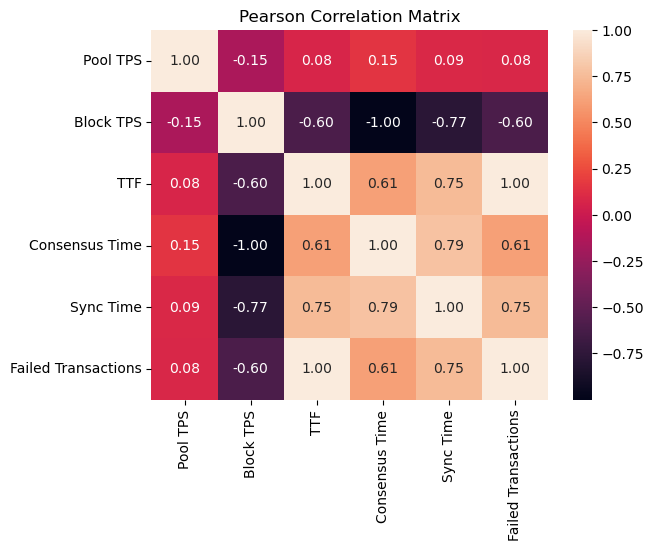}
\caption{Pearson correlation matrix of the averaged performance metrics across all deployments. Failed transactions drive synchronization and consensus time, which in turn reduce block throughput.}
\label{fig:correlation}
\end{figure}

Figure~\ref{fig:correlation} shows the Pearson correlation matrix of the averaged metrics. TTF, consensus time, synchronization time, and the number of failed transactions form a strongly positively correlated group, and the correlation between TTF and failed transactions is near unity: fraudulent transactions add to the load the pool must clear before a block can be committed. Block TPS falls as consensus time, synchronization time, and failed transactions rise, the expected consequence of a longer round on committed throughput.

\subsection{Bottlenecks}

Transaction pooling and processing dominate the round. Roughly \SI{83}{\percent} of the total time is spent there, and the pool processing rate is about \SI{60}{\percent} slower than the block creation and consensus rate. A large part of this gap follows from the single-laptop deployment, which cannot process transactions in parallel. Within consensus, the pre-prepare and prepare phases take about \SI{200}{\milli\second} each, jointly about \SI{84}{\percent} of consensus time, because both compute a Merkle root and hash the block; the commit phase averages \SI{13}{\milli\second}. The single most consequential risk we observe is that TTF rises to four to six seconds when most transactions are fraudulent, a threefold load increase. This state is rare under normal traffic, but a denial-of-service attack against the submission endpoint would produce it, so we flag it as a vulnerability to be addressed before deployment.

\subsection{Benchmark Comparison}

Table~\ref{tab:benchmark} places the system against common blockchains. The architecture reaches \num{1030} transactions per second with a TTF of about \SI{2.2}{\second}. For context, Visa processes on the order of \num{1667} transactions per second globally and PayPal on the order of \num{193} \cite{visa2023annual,paypal2023annual}. The system approaches Visa's operational rate and exceeds PayPal's, on a single laptop. Against other chains, it outperforms Bitcoin and Ethereum 1.0 on both throughput and finality, and is comparable to Ripple and SKALE, which trade finality against throughput differently. The comparison is drawn from vendor and third-party figures that are not measured on identical hardware, so it is indicative rather than controlled. The other systems are measured under their own transaction models and confirmation rules, and several are quoted at peak rather than sustained load, so the table establishes feasibility rather than a controlled ranking.

\begin{table}[!htbp]
\centering
\caption{Throughput and time to finality against common blockchain systems. Figures for other systems are vendor or third-party reports, not measured on identical hardware.}
\label{tab:benchmark}
\small
\begin{tabular*}{\textwidth}{@{\extracolsep{\fill}}l S[table-format=4.0] l@{}}
\toprule
\textbf{Architecture} & {\textbf{Block throughput (tx/s)}} & \textbf{Time to finality} \\
\midrule
This work      & 1030    & {\SI{2.2}{\second}} \\
Bitcoin        & 7       & {about \SI{1}{\hour}} \\
Ethereum 1.0   & {15--25} & {about \SI{6}{\minute}} \\
Ripple (XRP)   & 1500    & {about \SI{4}{\second}} \\
SKALE Network  & {400--700} & {about \SI{1.5}{\second}} \\
\bottomrule
\end{tabular*}
\end{table}

\subsection{Summary of Findings}

Three findings stand out. The architecture scales, since consensus metrics move by less than about \SI{2}{\percent} across the single-node, three-node, and four-node deployments, which is what the linear link count of Proposition~\ref{prop:complexity} predicts. The communication layer is not the bottleneck, because synchronization accounts for a small fraction of consensus time and Kafka carries the inter-node traffic without the delay a peer-to-peer design would add. The residual cost sits in transaction preparation rather than in consensus, which is a limit of this deployment rather than of the design. Equally clear is what the study does not establish: the security, governance, and incentive mechanisms are argued from the architecture rather than measured.

\section{Conclusion}
\label{sec:conclusion}

This paper presented a hybrid blockchain architecture that separates a decentralized consensus protocol, open to public validators, from a centralized network protocol governed by a single master node. Validators communicate through Kafka rather than over direct links, and every role runs as an independent containerized service on Kubernetes. The design fixes the level of centralization to a known constant, removes the direct peer-to-peer channels that Sybil and injection attacks use, and lets the two layers scale independently; Proposition~\ref{prop:complexity} shows the link count grows linearly with the number of validators instead of quadratically. The deployed prototype reaches a median block throughput of \num{1031.83} transactions per second and a median time to finality of \SI{2.20}{\second} on consumer-grade hardware, with consensus metrics changing by less than about \SI{2}{\percent} across single-node, three-node, and four-node deployments. The system is a proof of concept: transaction pooling, not consensus, is the dominant cost, and time to finality degrades under a flood of fraudulent transactions.

Future work follows in two directions. The implementation should be re-run on a dedicated cluster with true parallelism and hardware resource isolation, which separates the architectural limits from the limits of the test device and reveals where added validators stop helping, and the untested properties should be evaluated directly through simulated Sybil and majority-stake attacks and recovery tests. The architecture also admits a variation worth studying: a fully decentralized network layer in which the two protocols are governed by separate entities with different incentives, which raises the question of whether such a split preserves honesty without a fixed central authority.

\clearpage

\bibliographystyle{unsrt}
\bibliography{bibliography/blockchain}

\clearpage

\appendix

\section{Discussion}
\label{app:discussion}

The results support the central design decision, which is to split the chain into a decentralized consensus layer and a centralized network layer. Making the network layer the only centralized component fixes the level of centralization to a value every participant can name in advance. In Proof-of-Work and Proof-of-Stake the corresponding quantity emerges from the distribution of hash rate or stake and can drift without warning, at which point reversing it is difficult. Here the master node and the brokers are the whole of the centralized surface, and the master cannot alter block content or vote, because it neither signs nor validates.

The measurements also show where the design pays off and where it does not. Kafka carries the inter-node traffic at a synchronization cost of \SI{43}{\milli\second} out of \SI{496}{\milli\second} of consensus time, so the middleware is not the source of round delay. Most of the round is spent in the pool, which the asynchronous single-laptop deployment cannot run in parallel. The commit-phase cost of moving from one node to three, a \SI{77}{\percent} rise in a quantity that averages \SI{13}{\milli\second}, illustrates the same point in miniature: the relative change is large, the absolute change is small, and neither touches the round total.

\paragraph{Comparison with deployed systems.} Against the reported figures in Table~\ref{tab:benchmark}, the architecture sits between PayPal's operational rate and Visa's, and is comparable to Ripple and SKALE while balancing finality against throughput rather than trading one for the other. The comparison is drawn from vendor and third-party sources on different hardware, so it establishes feasibility rather than superiority. The strongest single claim we can make is that the block throughput exceeds \num{1000} transactions per second on a laptop with 10 CPU threads, without dedicated accelerators and without parallel transaction processing.

\paragraph{Design lessons.} Two implementation lessons carry beyond this system. First, when services are developed independently as microservices, task ordering must be designed before the services are written. The deadlock we hit came from phases advancing before their inputs existed, and it was resolved only by introducing the automaton in Figure~\ref{fig:automaton}; the automaton was cheaper to build once and reuse than to patch into each service after the fact. Second, a service should be validated natively at each level before it is deployed on Kubernetes: first on its own, then in a single-node deployment that exercises the middleware, then in a multi-node deployment that exercises synchronization. Failures were far easier to localize at the native levels than inside the orchestrator.

\paragraph{Threats to validity.} Three limits bound the strength of the results. The deployment runs on consumer hardware without resource isolation or true parallelism, so the measured throughput is a lower bound and the measured variability is an upper bound on what a dedicated cluster would produce. The frame is a local cluster, so inter-node latency does not reflect geographically distributed links. All nodes are assumed to be homogeneous, and transactions are generated uniformly in fixed batches, neither of which holds for real traffic. These limits are enumerated in full in Appendix~\ref{app:limitations}.

\section{System Model, Notation, and Threat Model}
\label{app:model}

\paragraph{Notation.} Table~\ref{tab:notation} collects the symbols used in the paper.

\begin{table}[!htbp]
\centering
\caption{Symbols used throughout the paper.}
\label{tab:notation}
\small
\begin{tabular*}{\textwidth}{@{\extracolsep{\fill}}l l@{}}
\toprule
\textbf{Symbol} & \textbf{Meaning} \\
\midrule
$n$ & number of validator nodes \\
$b$ & number of Kafka brokers \\
$s_i$ & stake held by validator $i$ \\
$f$ & number of Byzantine validators tolerated by the consensus protocol \\
$B$ & block size, in transactions \\
\bottomrule
\end{tabular*}
\end{table}

\paragraph{Trust model.} Assumption~\ref{ass:trust} states the model used in the analysis. The master node and the brokers are honest but may fail. The consensus protocol tolerates up to $f$ Byzantine validators. Clients are untrusted and can submit arbitrary or fraudulent transactions. The honest master is the sole authority over validator composition.

\paragraph{Adversary capabilities.} We consider three adversary types. A Sybil adversary creates many fake validator identities to acquire voting weight; the stake requirement makes this economically expensive, and admission through the master requires authentication, so identities cannot be created for free \cite{douceur2002sybil}. A majority adversary acquires enough stake to control consensus; the master can refuse admission to a validator it identifies as a threat, which interrupts a forming monopoly, and policies such as validator rotation and reward redistribution reduce the persistence of any majority. A host adversary compromises a validator through a host-level attack; because validators hold no direct links, the master can cut the compromised node from the topic and contain the damage, and the microservice deployment confines the failure to one service.

\paragraph{Attack surface.} Table~\ref{tab:surface} maps the principal threats to the mechanisms that address them. The mechanisms are architectural; none of them has been validated by simulation in this work, and we mark that status explicitly in Appendix~\ref{app:limitations}.

\begin{table}[!htbp]
\centering
\caption{Principal threats, the mechanisms that address them, and the strength of the current evidence.}
\label{tab:surface}
\small
\begin{tabular*}{\textwidth}{@{\extracolsep{\fill}}l p{0.28\textwidth} p{0.24\textwidth}@{}}
\toprule
\textbf{Threat} & \textbf{Mechanism} & \textbf{Status} \\
\midrule
Sybil identities & Stake-weighted admission and authenticated access to the topic & argued, not simulated \\
Majority stake control & Master-side denial of admission, validator rotation, reward redistribution & argued, not simulated \\
Node compromise & No direct links; master can isolate the node; microservice fault isolation & argued, not simulated \\
Message injection & All traffic passes through an authenticated Kafka topic & argued, not simulated \\
Middleware failure & Kafka replication and failover; validators can fork and reconcile & partially observed \\
\bottomrule
\end{tabular*}
\end{table}

\section{Scalability Analysis}
\label{app:scal}

Proposition~\ref{prop:complexity} gives the closed-form link counts. Two details are worth stating. First, the validator-to-broker term is $n b$ rather than $n$, but $b$ is a deployment constant chosen by the hosting organization, so the term is linear in $n$ for any fixed broker count. Second, the broker-to-broker term is $O(b)$ and does not grow with $n$. The total is therefore $O(n + b)$.

The crossover in Figure~\ref{fig:complexity} follows directly. The peer-to-peer count is $\binom{n}{2}$ and the proposed count is $nb + \binom{b}{2}$, both treated as connection counts. With $b = 3$ the peer-to-peer count is smaller than the proposed count only for very small $n$; once $n$ exceeds $b$, the proposed count stays below the quadratic curve and the gap widens. Independence between the layers is a separate scalability property: because consensus and networking run as distinct services, the number of validators can be increased by admitting more stake holders without changing the broker configuration, and the brokers can be scaled horizontally without changing consensus. The only constraint is that the two layers remain compatible, since validators that produce load faster than the network can carry it will block on the topic.

\section{Security Arguments and Economic Model}
\label{app:security}

\paragraph{Security.} The security argument rests on moving the attacker's cost from message volume to stake and authentication. In a peer-to-peer network a Sybil attacker pays only the cost of opening connections, which is why the threat is acute in open networks \cite{douceur2002sybil}. Here each identity must hold stake and must be admitted by the master, so inflating the validator count is expensive before any consensus influence is gained. The same admission control answers the majority-stake threat: a validator that approaches a controlling share can be denied participation, and the master can keep the draw weighted so that no faction reliably controls a round. Host compromise is bounded by the absence of direct links and by the master's ability to revoke topic access, and the containerized deployment from Section~\ref{sec:method} confines a compromised service to its own pod. Image integrity can be enforced by signing container images and verifying their digests before deployment. None of these mechanisms has been tested against a live adversary in this work.

\paragraph{Economic model.} The infrastructure splits into a compute side and a network side, and the split lowers the cost of scaling for the hosting organization. Validators carry the compute load and need only connect to the broker set, so their cost is dominated by computation, while the brokers carry the network load and scale horizontally. In a permissioned deployment the hosting organization funds the whole distributed infrastructure; here the compute side is funded by the validators, who are drawn in by the stake reward. A validator earns a share of transaction fees from the rounds it participates in, and the hosting organization holds the deposited stakes, so it can earn the time value of that capital. The model assumes that a validator that has invested in physical infrastructure will keep its stake deposited over the long term rather than withdrawing quickly; the rewards are structured to make that the rational choice, but this work does not test the assumption.

\section{Implementation Details}
\label{app:impl}

\paragraph{Repository layout.} The deployment is organized by role. The \texttt{node} service owns block creation, the Merkle tree, consensus, and networking; the \texttt{master} service owns stake handling and selection; the \texttt{transaction} service generates and submits load; and the \texttt{wallet} service manages keys and balances. Each service has its own manifest, ConfigMap, and container image, so a service can be rebuilt and redeployed on its own.

\paragraph{Deployment sequence.} Pods cannot be started simultaneously, because the node automaton expects the master and the brokers to be reachable before validators begin. The deployment sequence starts Zookeeper and Kafka first, then the master, then the transaction generator, and finally the validators, each stage gated on the readiness of the previous one.

\paragraph{Resource allocation.} Table~\ref{tab:resources} lists the minimum and maximum CPU and memory limits assigned to each Kubernetes service. The distribution keeps the cluster within the 10 CPU threads and \SI{30}{\giga\byte} of memory allotted to it while leaving headroom for the busiest service.

\begin{table}[!htbp]
\centering
\caption{Kubernetes services with their minimum and maximum resource requests.}
\label{tab:resources}
\small
\begin{tabular*}{\textwidth}{@{\extracolsep{\fill}}l l l@{}}
\toprule
\textbf{Service} & \textbf{Minimum limit} & \textbf{Maximum limit} \\
\midrule
Validator nodes            & 2 CPU threads, \SI{4}{\giga\byte} RAM & 3 CPU threads, \SI{6}{\giga\byte} RAM \\
Master node                & 1 CPU thread,  \SI{2}{\giga\byte} RAM & 2 CPU threads, \SI{4}{\giga\byte} RAM \\
Transaction simulator pod  & 1 CPU thread,  \SI{2}{\giga\byte} RAM & 2 CPU threads, \SI{4}{\giga\byte} RAM \\
Kafka pod                  & 1 CPU thread,  \SI{2}{\giga\byte} RAM & 2 CPU threads, \SI{4}{\giga\byte} RAM \\
Zookeeper pod              & 1 CPU thread,  \SI{2}{\giga\byte} RAM & 2 CPU threads, \SI{4}{\giga\byte} RAM \\
\bottomrule
\end{tabular*}
\end{table}

\paragraph{Kafka configuration.} Table~\ref{tab:kafka} lists the producer, consumer, and topic settings used for the evaluation, with the rationale for each. The configuration favours latency over throughput by accumulating small batches and acknowledging from one partition, and it uses a single broker, which is the minimum for a working cluster and does not exercise Kafka's fault tolerance.

\begin{table}[!htbp]
\centering
\caption{Kafka configuration used in the evaluation.}
\label{tab:kafka}
\small
\begin{tabular*}{\textwidth}{@{\extracolsep{\fill}}l p{0.42\textwidth} l@{}}
\toprule
\textbf{Setting} & \textbf{Purpose} & \textbf{Value} \\
\midrule
\texttt{linger.ms}         & Wait for messages to accumulate before sending & 10 \\
\texttt{min.batch.size}    & Minimum batch accumulated before send or receive & 64000 \\
\texttt{compression.type}  & Compression applied to messages on the wire & \texttt{lz4} \\
\texttt{acks}              & Acknowledgements required before a send completes & 1 \\
Brokers                    & Number of brokers, which sets fault tolerance & 1 \\
Topic partitions           & Partitions per topic, which sets transmission parallelism & 5 \\
\bottomrule
\end{tabular*}
\end{table}

\paragraph{Reproducibility.} The cluster is provisioned by applying the manifests under version control. Transaction batch size, block size, and the number of validators per round are set through ConfigMaps. Each node writes per-phase timings to a CSV file, which is extracted at the end of a run and analysed off-line. Raising the broker count and partition count in Table~\ref{tab:kafka} increases fault tolerance and transmission parallelism at the cost of the coordinating traffic they add.

\section{Extended Results}
\label{app:results}

\begin{figure}[!htbp]
\centering
\includegraphics[width=0.62\textwidth]{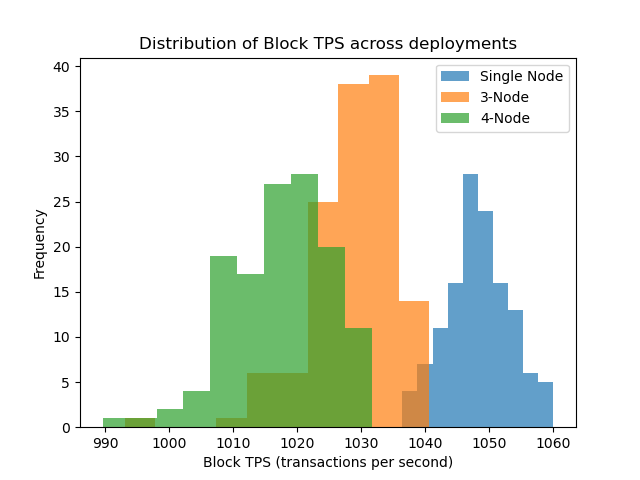}
\caption{Distribution of block TPS across the single-node, three-node, and four-node deployments.}
\label{fig:hist}
\end{figure}

\paragraph{Throughput distribution.} Figure~\ref{fig:hist} shows the distribution of block TPS across the three deployments. The median falls as validators are added, and the spread widens, with the four-node deployment showing the widest distribution. The small downward shift reproduces the trend in Table~\ref{tab:delta}; the widening reflects the growing uncertainty with which more nodes align their phase timing.

\begin{figure}[!htbp]
\centering
\includegraphics[width=0.62\textwidth]{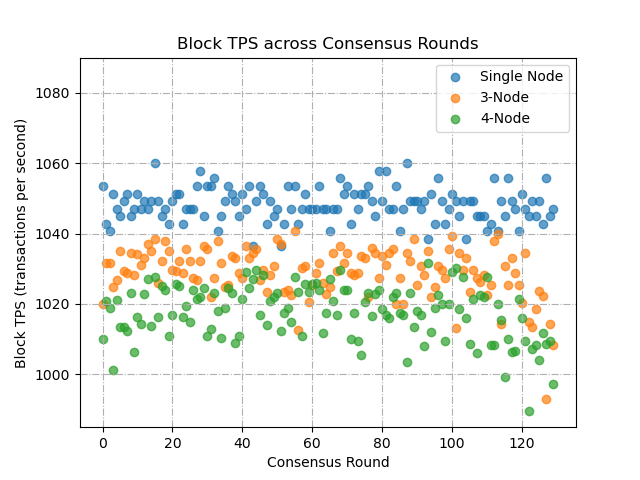}
\caption{Block TPS across consensus rounds for each deployment. Each series is stationary, and the deployments remain separated.}
\label{fig:rounds}
\end{figure}

\paragraph{Trend across rounds.} Figure~\ref{fig:rounds} plots block TPS against consensus round for each deployment. The distributions are stationary: throughput neither rises nor falls as rounds advance, which means throughput is independent of the round index even as the fraudulent-transaction load varies. The three deployments remain visibly separated, which matches the small but persistent gap in Table~\ref{tab:delta}.

\begin{figure}[!htbp]
\centering
\begin{subfigure}[t]{0.48\textwidth}
    \centering
    \includegraphics[width=\textwidth]{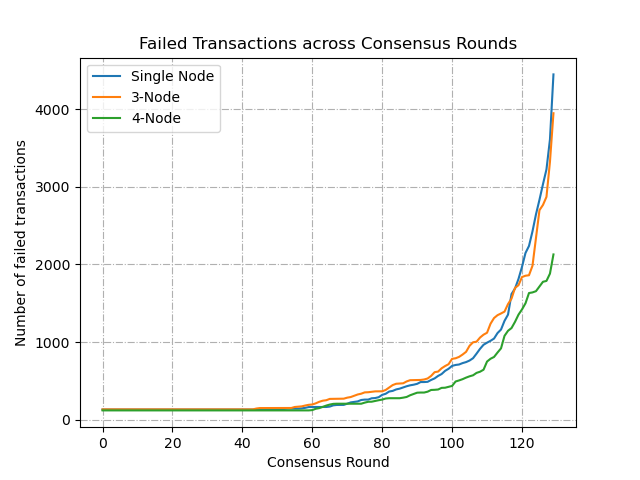}
    \caption{Failed transactions across consensus rounds.}
    \label{fig:failed}
\end{subfigure}
\hfill
\begin{subfigure}[t]{0.48\textwidth}
    \centering
    \includegraphics[width=\textwidth]{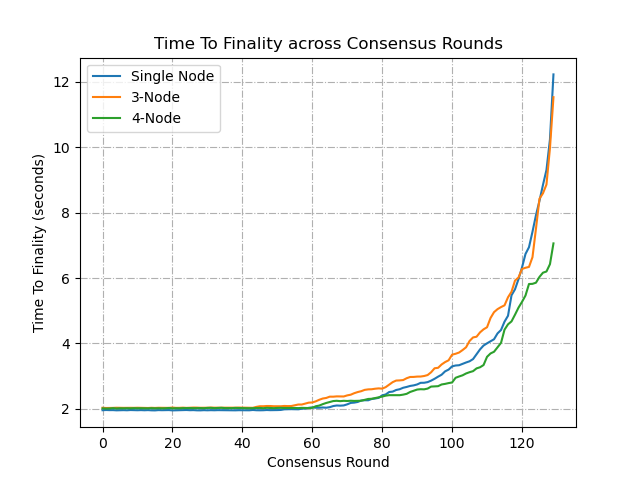}
    \caption{Time to finality across consensus rounds.}
    \label{fig:ttf}
\end{subfigure}
\caption{Finality under fraudulent load. Failed transactions and time to finality follow the same trend.}
\label{fig:finality}
\end{figure}

\paragraph{Finality under load.} Figures~\ref{fig:failed} and~\ref{fig:ttf} plot failed transactions and time to finality against round. The two curves move together: where the fraudulent-transaction count rises, the time to finality rises with it, which is the correlation reported in Section~\ref{sec:results} shown over the round sequence rather than as a coefficient. The relationship is close to exponential in the load, which is why a flood of fraudulent transactions is the single largest threat to finality observed in the study.

\section{Limitations and Assumptions}
\label{app:limitations}

We list the boundaries of the study explicitly.

\begin{enumerate}[leftmargin=*, nosep]
    \item \textbf{Hardware.} The cluster runs on a single consumer laptop without resource isolation or true parallelism, so the throughput figures are lower bounds and the variance figures are upper bounds relative to a dedicated cluster.
    \item \textbf{Network.} The deployment is local, so it does not capture bandwidth limits, geographic latency, or loss on wide-area links.
    \item \textbf{Middleware.} The Kafka configuration uses one broker and five partitions, which is a baseline rather than a tuned cluster, and it does not exercise failover.
    \item \textbf{Static resources.} Pod limits are fixed, so the deployment does not exercise autoscaling, which a production system would use under load.
    \item \textbf{Untested properties.} Security, fault tolerance, and the economic incentives are argued from architecture and are not measured. No Sybil, majority-stake, or broker-failure experiment was run.
    \item \textbf{Homogeneity.} All validators are assumed to share the same hardware and network capacity, and transactions are assumed to arrive uniformly in fixed batches. Neither holds for real traffic, which is bursty and heterogeneous.
\end{enumerate}

These limits define the natural next experiments: the same architecture on a dedicated, geographically distributed cluster, with adversarial security tests and a dynamic workload model.

\end{document}